\documentclass[graybox,vecphys]{svmult}

\usepackage{graphicx}        

\usepackage{newtxtext}       %
\usepackage{newtxmath}       

\makeindex             

\newcommand{\mR}{\mathbb{R}}
\newcommand{\cW}{\mathcal{W}}
\newcommand{\cT}{\mathcal{T}}
\newcommand{\cO}{\mathcal{O}}
\newcommand{\e}{\mathrm{e}}
\newcommand{\eps}{\epsilon}
\newcommand{\qexp}{\mathcal{E}\hspace{-.5pt}\mathit{xp}}
\newcommand{\qlog}{\mathcal{L}\hspace{-.3pt}\mathit{og}}
\newcommand{\abs}[1]{{\vert {#1} \vert}}
\newcommand{\bk}[1]{{\langle #1 \rangle}}
\DeclareMathOperator{\Op}{Op}
\DeclareMathOperator{\tr}{tr}
\DeclareMathOperator{\Tr}{Tr}
\DeclareMathOperator{\argmin}{argmin}

\begin{document}

\title*{Quantum drift-diffusion equations for a two-dimensional electron gas with spin-orbit interaction}
\titlerunning{Quantum drift-diffusion equations for a spin-orbit 2D electron gas}
%
%
\author{Luigi Barletti, Philipp Holzinger, and Ansgar J\"ungel}
%
\institute{Luigi Barletti \at Dip.\ di Matematica e Informatica ``U. Dini'', Universit\`a di Firenze,
\and
Philipp Holzinger \at Institute of Analysis and Scientific Computing, TU Wien, 
\and
Ansgar J\"ungel \at Institute of Analysis and Scientific Computing, TU Wien.}
%
%
\maketitle

\abstract*{Quantum drift-diffusion equations are derived for a two-dimensional electron gas with spin-orbit interaction of Rashba type. The (formal) derivation is based on a non-standard application of the usual mathematical tools, such as Wigner transform, Moyal product expansion, and Chapman-Enskog expansion. The main peculiarity consists in the fact that a non-vanishing current is already carried by the leading-order term in the Chapman-Enskog expansion. To our knowledge, this is the first example of quantum drift-diffusion equations involving the full (four-component) spin structure. Indeed, previous models were either quantum bipolar (two-component) or full-spin semiclassical (i.e.\ at leading-order in the Moyal expansion).}

\abstract{Quantum drift-diffusion equations are derived for a two-dimensional electron gas with spin-orbit interaction of Rashba type. The (formal) derivation turns out to be a non-standard application of the usual mathematical tools, such as Wigner transform, Moyal product expansion and Chapman-Enskog expansion. The main peculiarity consists in the fact that a non-vanishing current is already carried by the leading-order term in the Chapman-Enskog expansion. To our knowledge, this is the first example of quantum drift-diffusion equations involving the full spin vector. Indeed, previous models were either quantum bipolar (involving only the spin projection
on a given axis) or full spin but semiclassical.}

\section{Introduction}
\label{sec:1}
Spintronics is an alternative to electronics, where the bit of information is carried by the spin polarization and not by the current \cite{Zutic02}.
Spintronics must not be confused with quantum computing: in the latter, both the information and its processing are based on a relatively small number of
spins and are completely subject to the laws of quantum mechanics; in the former, the spin carriers are a large population and only the polarization
is the result of an average of many single spins.
Also in the case of spintronics, each spin carrier is subject to the laws of quantum mechanics and, for an accurate simulation of the behaviour of a spintronic device, it is very important to include quantum mechanical effects in the mathematical models.
A systematic way to construct mathematical models of quantum fluids (diffusive or hydrodynamic) has been introduced by Degond, Ringhofer, and M\'ehats
in Refs.\ \cite{DMR05,DR03} (see also the exposition in \cite{J2009}).
Their strategy is based on the quantum mechanical version of the Maximum Entropy Principle (MEP), which basically says that the fluid-dynamical (macroscopic)
equations, derived from an underlying kinetic (microscopic) model, can be closed by assuming that the microscopic state is the most probable one
compatible with the observed macroscopic quantities (densities, currents, etc.).
In turn, the most probable state is the one that maximises a suitable entropy functional, dictated by the laws of statistical mechanics.
The quantum MEP (Q-MEP) can be formulated in the standard (operator-based) formalism of statistical quantum mechanics or in the phase-space formalism
due to Wigner \cite{ZachosEtAl05}.
The operator form is more general, to the extent that it can also be applied to Hamiltonians defined in bounded domains (while the Wigner formalism
is only suited to the whole-space case).
However, the Wigner framework, being a quasi-classical description, is more suited to the semiclassical expansion of the quantum model, resulting
in  ``classical equations'' with ``quantum corrections''.

Diffusive models of particles with spin, subject to spin-orbit interactions, have been previously derived in Refs.\ \cite{BM2010,EH2014,PN2011}.
In Ref.\ \cite{EH2014}, two kinds of models are considered: the bipolar one, where only the projection of the spin on a given axis is considered,
and the spin-vector one, where all the components of the spin vector are present.
Such models are ``semiclassical'', which means that the drift-diffusion equations are not the standard ones because  (of course) they
contain the spin components, but the models do not incorporate non-local effects, such as the Bohm potential \cite{J2009}.
This is because the postulated equilibrium state is a classical Maxwellian for each spin component, while non-local effects only arise from a
quantum equilibrium state.
Reference \cite{PN2011} is a generalisation of \cite{EH2014}, where a more detailed collision operator is considered, with spin-dependent scattering rates.

The first application of the Q-MEP to the case of particles with spin-orbit interaction is given in Ref.\ \cite{BM2010}.
There, a two-dimensional electron gas (2DEG) with spin-orbit interaction of Rashba type \cite{Zutic02} is considered and the Q-MEP is used
to derive bipolar quantum drift-diffusion equations (QDDE) for the spin polarisation in the direction perpendicular to the 2DEG plane.
The obtained model is then expanded semiclassically in order to obtain classical drift-diffusion equations for the density and polarisation with quantum corrections.

Few results are available related to the existence analysis of spin drift-
diffusion models. The bipolar model was investigated in \cite{Gli08,GlGa10}.
An existence result for a diffusion model for the spin accumulation with fixed
electron current but non-constant magnetization was proved in
\cite{GaWa07,PuGu10}. Matrix spin drift-diffusion models were analyzed in
\cite{HoJu20,JNS15} with constant precession axis and in \cite{Zam14} with non-constant
precession vector. Numerical simulations for this model can be found in \cite{CJS16}.
Assuming a mass- and spin-conserving relaxation mechanism,
two full-spin drift-diffusion models were derived and analyzed in \cite{ZaJu13},
including spin-orbit interactions.
These model, however, do not contain ``quantum correction'' terms.

In the present paper, we derive spin-vector QDDE for the same spin-orbit system as in \cite{BM2010}.
As remarked before, this means that the QDDE that we derive here involve all the components of the spin vector.
The paper is organised as follows.
In Section \ref{sec:2}, we introduce the Rashba Hamiltonian, describing the spin-orbit interaction of each electron  in the 2DEG.
Moreover, some basic concepts of the spinorial Wigner-Moyal formalism are recalled.
In Section \ref{sec:3}, we set up the model at the kinetic level,
consisting of an evolution equation for the matrix-valued Wigner function, endowed with a collisional term that describes the relaxation of the system
to an equilibrium Wigner function obtained by the Q-MEP.
The formal diffusive limit of the kinetic model is analysed in Section \ref{sec:3}, which leads to the spin-vector QDDE
(Eqs.\ \eqref{QDDE1}, \eqref{bkTg}, and \eqref{bkTTg}).
In order to test the consistency of the obtained equations, we consider the semiclassical limit of the QDDE and show that it is
in accordance with the semiclassical equations derived in \cite{EH2014}.
\section{Physical and mathematical background}
\label{sec:2}
Let us consider a population of electrons confined in a two-dimensional potential well, described by the coordinates $(x_1,x_2)$ and subject to a spin-orbit interaction of
Rashba type \cite{Zutic02}.
The Hamiltonian of each electron has therefore the form
\begin{equation*}
	H =
	\begin{pmatrix}
	-\frac{\hbar^2}{2m}\Delta & -i\hbar\alpha_R \left(\partial_{x_2} + i\partial_{x_1} \right) \\
	-i\hbar\alpha_R \left( \partial_{x_2} - i\partial_{x_1}\right) & -\frac{\hbar^2}{2m}\Delta
	\end{pmatrix},
\end{equation*}
where $\alpha_R$ is the Rashba constant and $m$ is the (effective) electron mass.
In terms of the Pauli matrices, we can write
\begin{equation}
\label{Hdef0}
 H = -\frac{\hbar^2}{2m}\Delta\,\sigma_0 -  i\hbar \alpha_R\left(\partial_{x_2} \sigma_1 - \partial_{x_1}\sigma_2\right),
\end{equation}
where
$$
 \sigma_0 = \begin{pmatrix} 1 & 0 \\ 0 & 1   \end{pmatrix},
\quad
 \sigma_1 = \begin{pmatrix} 0 & 1 \\ 1 & 0   \end{pmatrix},
\quad
 \sigma_2 = \begin{pmatrix} 0 &-i \\ i & 0   \end{pmatrix},
\quad
  \sigma_3 =  \begin{pmatrix} 1 & 0 \\ 0 &-1   \end{pmatrix}.
$$
In the following, we will extensively make use of the algebra of the Pauli matrices.
Each $2\times 2$ matrix-valued quantity $a\in\mathbb{C}^{2\times 2}$
can be decomposed in Pauli components according to
$$
   a = \sum_{j=0}^3 a_j \sigma_j  = a_0\sigma_0 + \vec a \cdot \vec\sigma,
$$
where $\vec a = (a_1,a_2,a_3)$, $\vec \sigma = (\sigma_1,\sigma_2,\sigma_3)$,
and the components $a_k$ ($k = 0,1,2,3$) are real if and only if $a$ is hermitian.
By using the well-known identity
$$
\sigma_i\sigma_j = i\eps_{ijk}\sigma_k + \delta_{ij}\sigma_0,
\qquad 1\leq i,j,k, \leq 3,
$$
(where $\eps_{ijk}$ and $\delta_{ij}$ are, respectively, the Levi-Civita and Kronecker symbols),
it is straightforward to prove the following relations, mapping the matrix algebra on the Pauli components:
\begin{align}
\label{trace}
\tr(a) &= 2 a_0,
\\
\label{product}
ab &= (a_0b_0 + \vec a \cdot \vec b)\sigma_0 + (a_0 \vec b + b_0 \vec a + i \vec a \times \vec b) \cdot \vec\sigma,
\\
\label{commutator}
ab - ba &=  i \vec a \times \vec b \cdot \vec\sigma.
\end{align}
The Hamiltonian \eqref{Hdef0} can be written more concisely as
\begin{equation}
\label{Hdef1}
 H =  -\frac{\hbar^2}{2m}\Delta\,\sigma_0 -  i\hbar  \alpha_R\nabla^\perp \cdot \vec \sigma
\end{equation}
with the notation
$$
  \nabla = (\partial_{x_1}, \partial_{x_2}, 0), \quad
  \nabla^\perp = \nabla \times \vec e_3  = (\partial_{x_2}, -\partial_{x_1}, 0), \quad
	\vec e_3 = (0,0,1).
$$

We now combine the matrix algebra with the Wigner-Moyal calculus.
The following definitions and properties hold for suitably smooth functions.
Let us recall the definition of the Wigner transform,  $\varrho \mapsto a$, of a function $\varrho = \varrho(x,y)$, $x\in\mR^d$, $y\in\mR^d$, into
a phase-space function $a = a(x,p)$,  $x\in\mR^d$, $p\in\mR^d$:
\begin{equation*}
 a(x,p) = \cW(\varrho)(x,p) = \int_{\mR^d}
 \varrho \left( x + \frac{\xi}{2},  x - \frac{\xi}{2} \right) \e^{-ip \cdot \xi/\hbar}  d\xi
\end{equation*}
(see also Ref.\ \cite{ZachosEtAl05}).
We remark that, in our framework, we have $d= 2$, and the Wigner transform acts on
the matrix-valued functions $\varrho$ and $a$ componentwise.
The Wigner transformation is closely related to the Weyl quantization, $a \mapsto A$, that maps the phase-space function $a$ to
an operator $A$, according to
\begin{align*}
  (A\psi)(x) &= \left[ \Op(a)\psi \right](x)
  \\
  &=  \frac{1}{(2\pi\hbar)^d} \int_{\mR^{2d}}
  a\left( \frac{x+y}{2}, p \right)\,\psi(y)\,\e^{i(x-y)\cdot p/\hbar}\,dy\,dp.
\end{align*}
In the correspondence $A = \Op(a)$, the phase space function $a$ is often called the ``symbol'' of $A$.
\par
\begin{figure}[ht]
\sidecaption[t]
\includegraphics[scale=0.85]{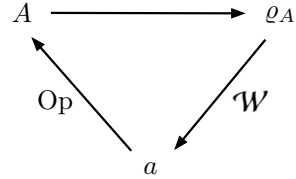}
\caption{The Wigner-Weyl correspondence: $A = \Op(a)$ is the operator associated to the phase-space function $a$, $\varrho_A$ is the
integral kernel of $A$, and $a = \cW(\varrho_A)$ is the Wigner transform of $\varrho_A$.}
\label{fig:1}
\end{figure}

The Wigner transform is the inverse of the Weyl quantization if one identifies the operator $A$ with its integral kernel $\varrho_A$.
In fact,
$$
  (A\psi)(x) = \int_{\mR^{d}}  \varrho_A(x,y)\,\psi(y)\,dy =  \int_{\mR^{d}}  \cW^{-1}(a)(x,y)\,\psi(y)\,dy.
$$
The Wigner-Weyl correspondence is summarized in Figure \ref{fig:1}.

The operator algebra is transferred to phase-space functions by the Wigner-Weyl correspondence.
In particular, the operator product gives rise to the definition of the Moyal product
$a\# b = \Op^{-1}(AB)$, where $A= \Op(a)$ and $B= \Op(b)$.
The Moyal product has an explicit expansion in powers of $\hbar$,
\begin{equation}
\label{MoyalExpansion}
a \# b = \sum_{k=0}^\infty \hbar^k a \#_k b,
\end{equation}
where
\begin{equation*}
 a \#_k b =  \frac{1}{(2i)^k} \sum_{\abs{\alpha} + \abs{\beta} = k}
\frac{(-1)^{\abs{\alpha}} }{ \alpha!\, \beta! }
\left(\nabla_x^\alpha \nabla_p^\beta a \right)\left(\nabla_p^\alpha \nabla_x^\beta b \right).
\end{equation*}
At the leading order of the expansion, we find the ordinary product
$a\#_0b = ab$, while at the first order, it is related to the Poisson bracket,
\begin{equation*}
a\#_1b = \frac{i}{2}\sum_{j = 1}^2\left(\partial_{x_j} a \, \partial_{p_j}  b -  \partial_{p_j} a \,\partial_{x_j}  b\right).
\end{equation*}

The operator trace $\Tr$ is equivalent to the integral on the phase-space of the matrix trace $\tr$ of its symbol, i.e.
\begin{equation*}
\Tr(A) = \int_{\mR^{2d}} \tr(a)(x,p)\,dx\,dp.
\end{equation*}
In particular, if $A$ represents some physical observable and $S$ represents the state of the system, and if $a = \Op^{-1}(A)$ and $w = \Op^{-1}(S)$
are the corresponding phase-space functions ($w$ is called the  {\em Wigner function} of the system), then the expected
value of the observable $A$ in the state $S = \Op(w)$ is
\begin{equation*}
\Tr(AS) = \int_{\mR^{2d}} \tr(aw)(x,p)\,dx\,dp.
\end{equation*}
By expressing this identity in terms of Pauli components (by using \eqref{trace} and \eqref{product}), we obtain the fundamental formula for the
expected values:
\begin{equation*}
\frac{1}{2}\Tr(AS) = \int_{\mR^{2d}} (a_0w_0 + \vec a \cdot \vec w) (x,p)\,dx\,dp.
\end{equation*}
This relation suggests to define the {\em local density} $n_A$ of the observable $A$ as
\begin{equation*}
n_A(x) = \int_{\mR^{d}} (a_0w_0 + \vec a \cdot \vec w) (x,p)\,dp = \bk{n_A} (x),
\end{equation*}
where we introduced the notation
$\bk{f} = \int_{\mR^{d}} f \,dp$.
Since our goal is to derive a spinorial diffusive model, the local densities we are interested in are the position
density $n_0$ (observable $\frac{1}{2}\sigma_0$) and  the spin density $\vec n$ (observable $\frac{1}{2}\vec \sigma$),
given by
\begin{equation*}
 n_0(x) = \int_{\mR^{2d}} w_0 (x,p)\,dp, \qquad \vec n(x) = \int_{\mR^{2d}} \vec w(x,p)\,dp.
\end{equation*}
We remark that an operator $S$ representing the state of a quantum system must be a positive operator with unit trace.
In particular, $(S\psi)(x)$ is a positive definite matrix for all two-component wave functions $\psi$ and for a.e.\ $x$.
This fact leads to constraints on the functions $n_k$, $k = 0,1,2,3$, namely $n_0 \geq 0$ and $n_1,n_2,n_3 \in \mR$ with
$$
  n_1^2 + n_2^2 + n_3^2 \leq n_0^2
$$
(for a.e.\ $x$).
If $n_1^2 + n_2^2 + n_3^2 = n_0^2$, then $S$ and $w = \Op^{-1}(S)$ represent a pure state while if
$n_1^2 + n_2^2 + n_3^2 < n_0^2$, then $S$ and $w$ represent a ``mixed'' (statistical) state.
\section{Transport picture}
\label{sec:3}
We shall now derive a mesoscopic-level (kinetic) transport model for our two-dimensional electron gas.
\subsection{Transport equation}
Let $S(t)$ be the time-dependent density operator, representing the statistical quantum mechanical state at time $t$,
let $\varrho(x,y,t)$ be the associated density matrix (i.e.\ the integral kernel of $S(t)$) and $w(x,p,t) = \cW(\varrho)$
the corresponding Wigner function.
The evolution equation for $S(t)$ is the statistical version of the Schr\"odinger equation, that is the Von  Neumann equation
\begin{equation*}
  i\hbar \partial_t S = (H+V) S - S(H+V),
\end{equation*}
where $H$ is the Rashba Hamiltonian \eqref{Hdef1} and $V = V(x)\sigma_0$ represents an external electrostatic potential (e.g.\ a gate potential).
In terms of the density matrix, this equation reads as follows:
\begin{equation*}
  i\hbar \partial_t \varrho =\left(  -\frac{\hbar^2}{2m}\left(\Delta_x  - \Delta_y\right) + V(x)  - V(y) \right) \varrho
  -i\hbar\alpha_R \big(\nabla_x^\perp\cdot\vec\sigma\varrho - \nabla_y^\perp  \varrho \cdot\vec\sigma \big)
\end{equation*}
The evolution equation for the Wigner function $w$ is obtained by applying the Wigner transformation to both sides of
the last equation.
This results in
\begin{equation*}
  i\hbar \partial_t w = \left\{ h+V ,  w\right\}_\#,
\end{equation*}
where
\begin{equation*}
  h(x,p) =  \frac{\abs{p}^2}{2m} \sigma_0 + \alpha_R p^\perp \cdot \vec \sigma
\end{equation*}
is the symbol of the Rashba Hamiltonian (as usual $p^\perp = p \times \vec e_3 = (p_2,-p_1,0)$)
and $\{\cdot,\cdot\}_\#$ is the Moyal bracket
\begin{equation*}
  \{a,b \}_\# =  a\# b - b\#a.
\end{equation*}
By explicitly computing this bracket and decomposing the matrix equation in the Pauli components, we obtain the following system
for the trace and spin parts of $w$:
\begin{equation}
\label{WE1}
\begin{aligned}
  &\partial_t w_0 = -\frac{1}{m}p\cdot \nabla_x w_0 - \alpha_R \nabla_x^\perp \cdot \vec w + \Theta_\hbar[V] w_0,
\\
  &\partial_t \vec w = -\frac{1}{m}p\cdot \nabla_x \vec w - \alpha_R \nabla_x^\perp w_0  + \Theta_\hbar[V] \vec w
  + \frac{2\alpha_R}{\hbar} p^\perp \times \vec w,
\end{aligned}
\end{equation}
where
\begin{align}
\label{Thetadef}
\Theta_\hbar[f]  &= \frac{1}{i\hbar}\left[ f\left(x + \frac{i\hbar}{2}\nabla_p \right) - f\left(x - \frac{i\hbar}{2}\nabla_p \right) \right]
\nonumber \\
&= \sum_{j=0}^\infty (-1)^j \left(\frac{\hbar}{2}\right)^{2j} \sum_{\abs{\alpha} = 2j+1} \frac{1}{\alpha!} \nabla_x^\alpha f \nabla_p^\alpha
\end{align}
is the usual force term of the Wigner equation \cite{BFM2014,J2009,ZachosEtAl05}.
Note that the leading order term of the last expansion corresponds to the force term in the classical transport equation, namely
$$
  \Theta_\hbar[V] \xrightarrow{\hbar \to 0} \nabla_x V \cdot \nabla_p.
$$

In order to study the diffusion asymptotics of our system, the purely hamiltonian dynamics described by Eq.\ \eqref{WE1} must be
supplemented with a collisional mechanism.
If we want to remain in a rigorous quantum-mechanical setting, we cannot expect to adopt a detailed description of collisions.
However, since our goal is to obtain the diffusive limit of our model, only very general properties of the collision dynamics are needed,
such as conservation properties.
Therefore, the optimal strategy to insert a relatively simple collisional mechanism, and to respect at the same time the rules of quantum
mechanics, is to adopt a relaxation-time term making the system relax to a suitable quantum equilibrium state
\cite{Arnold96,DMR05,DR03,J2009}.
We therefore re-write  Eq.\ \eqref{WE1} with suitable relaxation-time terms:
\begin{equation}
\label{WBGK}
\begin{aligned}
  &\partial_t w_0 = -\frac{1}{m}p\cdot \nabla_x w_0 - \alpha_R \nabla_x^\perp \cdot \vec w + \Theta_\hbar[V] w_0 + \frac{1}{\tau_p}\left(g_0 - w_0  \right)
\\
  &\partial_t \vec w = -\frac{1}{m}p\cdot \nabla_x \vec w - \alpha_R \nabla_x^\perp w_0 +  \Theta_\hbar[V] \vec w + \frac{2\alpha_R}{\hbar} p^\perp \times \vec w
    + \frac{1}{\tau_p}\left(\vec g - \vec w  \right)
\end{aligned}
\end{equation}
where $g = g_0\sigma_0 + \vec g \cdot \vec \sigma$ is the equilibrium Wigner function that will be specified later on.

Before that, and in view of the diffusion asymptotics, let us rewrite Eq.\ \eqref{WBGK} in a non-dimensional form.
Let $T_0$ be the (given) temperature of the thermal bath with which our electron population is assumed to be in equilibrium.
The reference energy $E_0$ is taken as the thermal energy, given by
$$
k_B T_0=E_0,
$$
where $k_B$ denotes the Boltzmann constant.
The associated thermal momentum is
$$
 p_0 = \sqrt{mk_BT_0}.
$$
Let us also fix a reference length $x_0$ (e.g., the device size) and take the reference time $t_0$ as
$$
  t_0 = \frac{mx_0}{p_0},
$$
which is the time it takes an electron, traveling at the reference thermal velocity, to cross the reference length.
Then, in Eq.\ \eqref{WBGK} we switch to non-dimensional variables with the substitutions
$$
x \to x_0 x, \qquad  t\to t_0 t, \qquad  p \to p_0p,  \qquad  V \to E_0V
$$
(for the sake of simplicity, the new non-dimensional variables are denoted by the same symbols as the dimensional ones).
We obtain in this way
\begin{equation}
\label{WBGKscaled}
\begin{aligned}
  &\partial_t w_0 = - p\cdot \nabla_x w_0 - \eps\alpha \nabla_x^\perp \cdot \vec w + \Theta_\eps[V] w_0 + \frac{1}{\tau}\left(g_0 - w_0  \right),
\\
  &\partial_t \vec w = - p\cdot \nabla_x \vec w - \eps \alpha \nabla_x^\perp w_0 + \Theta_\eps[V] \vec w + 2\alpha p^\perp \times \vec w
  + \frac{1}{\tau}\left(\vec g - \vec w  \right) .
\end{aligned}
\end{equation}
Here, two important non-dimensional parameters have been introduced,
\begin{equation*}
\eps=\frac{\hbar}{x_0p_0},\qquad \tau = \frac{\tau_p}{t_0}.
\end{equation*}
The ``semi-classical'' parameter $\eps$ is the scaled Planck constant: roughly speaking, the smaller it is, the further we zoom out from the quantum scale and approach the classical scale.
The diffusive parameter $\tau$ is the scaled collision time: the smaller it is, the more collisions occur in the reference time, making the diffusive
regime predominate on the ``ballistic'' one.
Moreover,
\begin{equation*}
\alpha = \frac{m x_0 \alpha_R}{\hbar}
\end{equation*}
is the non-dimensional Rashba constant.
Since $\eps \alpha = m \alpha_R/p_0$, we see that $\alpha$ is the coefficient of proportionality between $\eps$  and the ratio of $\alpha_R$
(which has the physical dimension of a velocity) and the thermal velocity $p_0/m$.
This choice makes the Rashba constant scale with $\eps$ and gives the correct result in the semiclassical limit $\eps \to 0$
(see Section \ref{sec:4.3} and Ref.\ \cite{BM2010}).
\subsection{Maximum entropy principle}
We now come to the description of the quantum equilibrium function appearing in the transport equation \eqref{WBGKscaled}.
According to the theory developed in Refs.\ \cite{DMR05,DR03}  (see also \cite{BFM2014,J2009}), we choose the equilibrium
Wigner function $g = g_0\sigma_0 + \vec g \cdot \vec \sigma$ as the minimiser of a suitable quantum entropy-like functional, with the
constraint of positivity and fixed densities, which is the quantum version of the well-known Maximum Entropy Principle.
Physically speaking, this means that $g$ is assumed to be the most probable microscopic state compatible with the observed
macroscopic density.
This is rigorously expressed in our case as follows.
\par
\medskip
\noindent
{\bf Quantum Maximum Entropy Principle (Q-MEP).}
{\it Let $n = n_0\sigma_0 + \vec n \cdot \vec \sigma$ be a given matrix-valued function of $x$ and $t$,
with
$$
n_0 > 0, \qquad  n_1,n_2,n_3 \in \mR, \qquad  n_1^2 + n_2^2 + n_3^2 < n_0^2,
$$
for a.e.\ $x\in\mR^2$ and $t >0$.
The equilibrium Wigner function $g$ is given by
\begin{equation*}
g = \argmin \left\{  \mathcal{H}(w) \mid \Op(w) > 0, \ \bk{w} = n \right\},
\end{equation*}
where $\mathcal{H}$ is the quantum free-energy functional given (in non-dimensional variables) by
\begin{equation}
\label{EntropyFunctional}
 \mathcal{H}(w) = \frac{1}{2} \tr\left( \int_{\mR^6} \left( w\qlog(w) - w + h\# w \right)(x,p)\, dxdp \right)
\end{equation}
and  $\qlog$ is the ``quantum logarithm'' defined as
\begin{equation*}
\qlog(w) = \cW \left( \log(\Op(w)) \right)
\end{equation*}
($\log$ being the operator logarithm).}
\par
\medskip
Note that the constraints on $n$ are consistent with the requirement that $w$ represents a quantum mixed state, according to the remark at the end of Sec.\ \ref{sec:2} (see also \cite{MP10,MP11}).
\par
\smallskip
Then, $g$ is defined as the Wigner function that minimises the quantum entropy (or, more precisely, the free energy, which is the energy minus the entropy)
under the constraint of the given density.
Note that the condition $\Op(g) > 0$ means that $g$ must be a genuine Wigner function (i.e.\ the Wigner transform of a density operator).
The entropy functional \eqref{EntropyFunctional} is the phase-space equivalent of the Von Neumann entropy (free energy, more precisely): if $S = \Op(w)$
is the density operator, then
$$
   \mathcal{H}(w) = \Tr\left( S\log(S) - S +  HS\right).
$$
A formal proof of the following theorem makes use of the mathematical techniques adopted  in similar contexts (see, e.g., Ref.\ \cite{BF2010});
however the application of these techniques to the full-spin case is far from being straightforward and a detailed proof is deferred to a forthcoming paper.
Rigorous proofs also exist, but only for the simpler case of a one-dimensional system of scalar
(non-spinorial) particles in an interval with periodic boundary conditions, see Refs.\ \cite{MP10,MP11}.
\par
\medskip
\noindent
{\bf Theorem.}
{\it The matrix-valued Wigner function $g$, satisfying the above constrained minimisation problem, exists and is given by
\begin{equation}
\label{gdef}
g = \qexp(-h + a), \qquad \bk{g} = n,
\end{equation}
where $a = a_0\sigma_0 + \vec a \cdot \vec \sigma$ is a matrix of Lagrange multipliers and
\begin{equation*}
\qexp(w) = \cW \left( \exp(\Op(w)) \right)
\end{equation*}
(with $\exp$ the operator exponential).
}
\par
\medskip
Our model is now completed by using $g$ given by \eqref{gdef} as the equilibrium function in the Wigner equation \eqref{WBGK}.
We remark that the quantum equilibrium function $g$ is quite a complicated object, it is a non-local function of the Lagrange multipliers,
which are implicitly related to the densities $n_0$ and $\vec n$ by the four integral constraints $\bk{g} = n$, i.e.\ $\bk{g_0} = n_0$ and $\bk{\vec g} = \vec n$.
However, it is possible to make the model more explicit by performing a semiclassical expansion of $g$, made possible by the semiclassical
expansion \eqref{MoyalExpansion} of the Moyal product.
\section{Diffusion picture}
\label{sec:4}
Let us now formally derive the diffusion asymptotics of the kinetic model introduced in the previous section.
\subsection{Chapman-Enskog expansion}
To shorten the notation, we denote by $\cT$ the transport operator
\begin{align*}
\cT w :=  \frac{1}{i\eps}\left\{h+V, w \right\}_\# &=  \left(- p\cdot \nabla_x w_0 - \eps\alpha \nabla_x^\perp \cdot \vec w + \Theta_\eps[V] w_0\right)\sigma_0
\\
&\phantom{ixx}+ \left( - p\cdot \nabla_x \vec w - \eps \alpha \nabla_x^\perp w_0 + \Theta_\eps[V] \vec w + 2\alpha p^\perp \times \vec w \right)\cdot\vec\sigma,
\end{align*}
so that the scaled Wigner equation \eqref{WBGKscaled} is concisely written as
\begin{equation}
\label{WBGKconcise}
\tau \partial_t w = \tau \cT w + g - w.
\end{equation}
The diffusion asymptotics is obtained by means of the Chapman-Enskog expansion \cite{Cercignani88,J2009}, by expanding the equation for the macroscopic density $n = \bk{w}$,
\begin{equation*}
  \partial_t n = \partial_t^{(0)} n + \tau \partial_t^{(1)} n + \tau^2 \partial_t^{(2)} n+ \cdots,
\end{equation*}
and the microscopic state,
\begin{equation}
\label{wexp}
  w = w^{(0)} + \tau w^{(1)} + \tau^2 w^{(2)} + \cdots .
\end{equation}
We remark that it is only the equation for $n$ that is expanded, and not $n$ itself, which is an $\cO(1)$ quantity with respect to $\tau$.

Integrating \eqref{WBGKconcise} with respect to $p$ and recalling that $\bk{g-w} = 0$ (which follows from \eqref{gdef} and reflects the conservation
of the number of particles and the spin in the collisions), we can identify the $k$-th order time derivative of $n$ by
\begin{equation*}
  \partial_t^{(k)} n = \bk{\cT w^{(k)}} .
\end{equation*}
To compute $w^{(k)}$, we substitute \eqref{wexp} in \eqref{WBGKconcise}.
This yields, at leading and at first order in $\tau$,
\begin{equation*}
w^{(0)} = g, \quad w^{(1)} = \cT g - \partial_t  g,
\end{equation*}
respectively. Therefore,
\begin{equation}\label{pt1}
  \partial_t^{(0)} n = \bk{\cT g}, \quad
  \partial_t^{(1)} n = \bk{\cT\cT g} - \bk{\cT  \partial_t g}.
\end{equation}

The function $g$ depends on time only through its (functional) dependence on $n$,  according to \eqref{gdef}.
Then, at the same order of approximation, we can also write
\begin{equation}
\label{ptg}
  \partial_t  g = \frac{\delta g}{\delta n} \circ \partial_t n \approx    \frac{\delta g}{\delta n} \circ \partial_t^{(0)} n
  =   \frac{\delta g}{\delta n} \circ \bk{\cT g} ,
\end{equation}
where $\circ$ denotes the componentwise product, resulting from the chain rule
$$
\frac{\delta g}{\delta n}  \circ \partial_t n \equiv \sum_{k=0}^3  \frac{\delta g}{\delta n_k} \partial_t n_k .
$$
Using \eqref{pt1} and \eqref{ptg} and neglecting higher-order terms, we obtain the quantum drift-diffusion (QDDE)
equation for $n$:
\begin{equation}
\label{QDDE1}
  \partial_t  n = \bk{\cT g} + \tau\bk{\cT\cT g} - \tau \bigg\langle\cT  \frac{\delta g}{\delta n} \bigg\rangle \circ \bk{\cT g} .
\end{equation}
We remark the following:
\begin{enumerate}
\item
The QDDE \eqref{QDDE1} is, formally, a closed equation for $n$, since $g$ depends on $n$ through \eqref{gdef}.
\item
The term $\tau\bk{\cT\cT g}$ is the truly diffusive term in the equation, to the extent that it is the only term that appears in the standard cases
(i.e.\ classical or quantum scalar particles \cite{DMR05,DR03,J2009}).
\item
The term $\bk{\cT g}$, which is equal to zero for standard particles, does not vanish for spin-orbit electrons (see below). This is the reason why we were forced to use a hydrodynamic scaling instead of the usual diffusive one.
As a consequence, the Chapman-Enskog procedure produces the additional terms $\bk{\cT g}$ and $\tau \bk{\cT  \frac{\delta g}{\delta n} } \circ \bk{\cT g}$
in the diffusive equations.
\end{enumerate}
The last point deserves some additional comments.
In the usual situation, the diffusion asymptotics is derived from the transport, or kinetic, equation in the so-called diffusive scaling, i.e.\ obtained
by a further rescaling of time, $t \mapsto t/\tau$.
This means that the system is observed on a very long time scale, in which the collision time is $\tau^2$ (the hydrodynamic scaling being instead the one in which
the collision time is $\tau$).
This is because in the standard case, if collisions do not conserve the momentum, one has $\bk{\cT g} = 0$, which reflects the fact that the equilibrium
state carries no current.
Therefore, a purely diffusive current manifests in the longer, diffusive, time scale.
In the present situation, even though the collisions do not conserve the momentum, $g$ still carries a current, that is due to the peculiar form of the spin-orbit
interaction.
This implies that a current, $\bk{\cT g}$, already appears at the hydrodynamic scale.
Moreover, at order $\tau$ the additional term $\tau \bk{\cT  \frac{\delta g}{\delta n} } \circ \bk{\cT g}$ appears.
A formally analogous term appears also in the derivation of the classical hydrodynamic equation: in that case it contains the viscosity \cite{Cercignani88}.
In the present context, its interpretation is not so clear.
We point out that the two non-standard terms $\bk{\cT g}$ and $\tau \bk{\cT  \frac{\delta g}{\delta n} } \circ \bk{\cT g}$ are ``small'' in a semiclassical perspective, because,
as we shall see later, they vanish at leading order in $\eps$.
\subsection{Quantum drift-diffusion equation}
In order to recast \eqref{QDDE1} in a more explicit form, note that we can write
$$
  \cT g = \frac{1}{i\eps}\left\{h+V, g \right\}_\# = \frac{1}{i\eps}\left\{h-a, g \right\}_\# + \frac{1}{i\eps}\left\{V+a, g \right\}_\#
  =  \frac{1}{i\eps}\left\{V+a, g \right\}_\#,
$$
where $a$ is the matrix of Lagrange multipliers; see \eqref{gdef}. In fact,
\begin{equation}
\label{commut}
  \left\{h-a, g \right\}_\# = 0,
\end{equation}
because $g = \qexp(-h + a)$ and therefore, \eqref{commut} is just the expression in the Wigner-Moyal formalism of the commutativity
of the operator $H-A$ with its exponential $\exp(-H+A)$.
Recalling that $V$ and $a$ do not depend on $p$, we find that
\begin{align}
\label{Tg}
\cT g &= \frac{1}{i\eps}\left\{V+a, g \right\}_\# =
  \left( \Theta_\eps[V+a_0]g_0 +  \Theta_\eps[\vec a]\cdot \vec g\right) \sigma_0
\\
&\phantom{xx}{}+  \left( \Theta_\eps[V+a_0]\vec g +  \Theta_\eps[\vec a] g_0 + \eps^{-1}  \Theta^+_\eps[\vec a] \times \vec g \right)\cdot \vec \sigma, \nonumber
\end{align}
where $\Theta_\eps$ is given by \eqref{Thetadef} and $\Theta^+_\eps$ is defined as follows:
\begin{align}
\label{Theta+def}
\Theta^+_\eps[f]  &= \frac{1}{i\eps}\left[ f\left(x + \frac{i\eps}{2}\nabla_p \right) + f\left(x - \frac{i\hbar}{2}\nabla_p \right) \right]
\\
&= \sum_{j=0}^\infty (-1)^j \left(\frac{\eps}{2}\right)^{2j} \sum_{\abs{\alpha} = 2j} \frac{1}{\alpha!} \nabla_x^\alpha f \nabla_p^\alpha . \nonumber
\end{align}

We infer from \eqref{Thetadef} (with $\eps$ instead of $\hbar$)
and \eqref{Theta+def} the following properties:
\begin{equation*}
  \bk{\Theta_\eps[f]w} = 0, \quad  \bk{p_j \Theta_\eps[f]w} = - \partial_{x_j} f \,\bk{w},
  \quad \bk{\Theta^+_\eps[f]w} = 2 f \bk{w}.
\end{equation*}
Then, recalling that $\bk{g} = n$,
\begin{equation}
\label{bkTg}
  \bk{\cT g} = 2\eps^{-1} \vec a \times \vec n \cdot \vec \sigma.
\end{equation}
This represents explicitly the above-mentioned residual spin-orbit current at equilibrium.
We see that a condition for this current to vanish is
\begin{equation}
\label{commcond}
  \vec a \times \vec n = 0,
\end{equation}
which is equivalent to the commutativity of the matrices $n$ and $a$ (see Eq.\  \eqref{commutator}).
This explains why in Ref.\ \cite{BM2010}, concerning the bipolar case, only the standard diffusion term $\bk{\cT \cT g}$
has been found: in that case the matrices $n$ and $a$ are both diagonal.

Now, for a generic $w$, we have
\begin{align}
\label{aux}
 \bk{\cT w} &= \left( -\partial_j \bk{p_j w_0} - \eps\alpha \nabla^\perp \cdot \bk{\vec w} \right)\sigma_0
 \\
 &\phantom{xx}{}+  \left( - \partial_j \bk{p_j \vec w} - \eps \alpha \nabla^\perp \bk{w_0} +  2\alpha \bk{p^\perp \times \vec w} \right)\cdot\vec\sigma \nonumber
\end{align}
(where $\partial_j \equiv \partial_{x_j}$ and  summation over $j = 1,2$ is assumed).
Substituting $w = \cT g$ in \eqref{aux}, where $\cT$ is defined in \eqref{Tg}, yields
\begin{align}
  \bk{\cT\cT g}  &=  \left\{   \partial_j \left[ n_0 \, \partial_j (V+a_0) + \vec n \cdot  \partial_j \vec a \right] - 2\alpha \nabla^\perp \cdot (\vec a \times \vec n) \right\} \sigma_0
\label{bkTTg} \\
  &\phantom{xx}{}+ \Big\{  \partial_j \left[ \vec n \, \partial_j (V+a_0) + n_0 \partial_j \vec a  - 2\eps^{-1} \vec a \times \bk{p_j \vec g}\right]
\nonumber \\
  &\phantom{xx}{}-2\alpha \left[ \nabla^\perp(V+a_0) \times \vec n + (\nabla^\perp  \times \vec a ) n_0
  - 2\eps^{-1} \alpha \bk{p^\perp \times(\vec a \times \vec g)}  \right]  \Big\} \cdot \vec \sigma . \nonumber
 \end{align}
Equations \eqref{bkTg} and \eqref{bkTTg} express the first and the second terms in the quantum drift-diffusion equations \eqref{QDDE1} in terms of the
Lagrange multipliers (no such explicit expression has been found for the third term).

We remark that the Lagrange multipliers depend on the densities $n$ via the constraint $\bk{g} = n$.
Even though this fact makes \eqref{QDDE1} a closed equation for $n$, nevertheless the dependence of $a$ on $n$ is
very implicit and non-local, since it comes from integral constraints on a quantum exponential, involving back and forth Wigner transforms.
Numerical methods to solve QDDE of this kind exist \cite{BMNP2005,GM2006}.
However, the optimal use of a QDDE is expanding it semiclassically (i.e. in powers of $\eps$), in order to obtain ``quantum corrections'' to classical QDD
\cite{BM2010,BF2010,DMR05,DR03,J2009}.
This will be the subject of a future work.
For the moment, we shall limit ourselves to consider the semiclassical limit $\eps \to 0$ of \eqref{QDDE1},
just to check if our model allows us to recover the semiclassical drift-diffusion equations for spin-orbit electrons already known in the literature \cite{EH2014}.
\subsection{Semiclassical limit}
\label{sec:4.3}
The semiclassical limit is obtained from the fully quantum model \eqref{QDDE1},
\eqref{bkTg}, and \eqref{bkTTg} by expanding $g$ and $a$ in powers
of $\eps$ and retaining only the terms of order $\cO(\eps^0)$.
This would require the expansions of $g$ and $a$ up to $\cO(\eps^1)$, because of the terms of order $\eps^{-1}$ appearing in \eqref{bkTg}
and \eqref{bkTTg}.
So it is easier to compute directly the right-hand side of \eqref{QDDE1}, neglecting
all terms of order $\eps$ and using the leading-order approximation of $g$,
that is
\begin{equation*}
 g(x,p,t) \approx \e^{-p^2/2} \e^{a(x,t)} = \frac{1}{2\pi} \e^{-p^2/2} n(x,t).
\end{equation*}
We remark that this is indeed the semiclassical equilibrium distribution (see, e.g., Ref.\ \cite{EH2014}). Within this approximation, we have
$\bk{\cT g} \approx 0$ (and then, of course, also $\bk{\cT  \frac{\delta g}{\delta n} } \circ \bk{\cT g} \approx 0$) as well as
\begin{align*}
  \bk{\cT\cT g}  &\approx  \partial_j \left(  \partial_j  n_0 +n_0\partial_j V \right) \sigma_0
\\
  &\phantom{xx}{}+ \Big\{  \partial_j \left[  \partial_j  \vec n +\vec n \partial_j V  + 4\alpha A_j(\vec n)  \right]
     -2\alpha \nabla^\perp V \times \vec n - 4\alpha^2 B( \vec n)  \Big\} \cdot \vec \sigma ,
 \end{align*}
where
\begin{equation*}
 A_1(\vec n) = \begin{pmatrix} -n_3 \\ 0 \\ n_1  \end{pmatrix},
 \qquad
 A_2(\vec n) = \begin{pmatrix} 0 \\ -n_3 \\ n_2  \end{pmatrix},
  \qquad
 B(\vec n) = \begin{pmatrix} n_1 \\  n_2 \\ 2n_3  \end{pmatrix}.
\end{equation*}
Then, as a leading-order approximation of the quantum drift-diffusion equations \eqref{QDDE1}, we arrive to
\begin{align*}
  \partial_t n_0 &= \partial_j \left(  \partial_j  n_0 +n_0\partial_j V \right),
\\
 \partial_t \vec n &= \partial_j \left[  \partial_j  \vec n +\vec n \partial_j V  + 4\alpha A_j(\vec n)  \right]
      -2\alpha \nabla^\perp V \times \vec n - 4\alpha^2 B( \vec n).
\end{align*}

The semiclassical drift-diffusion equations derived in Ref.\ \cite{EH2014} coincide with our equations in the case of constant relaxation time and
purely spin-orbit interaction field. (In Ref.\ \cite{EH2014} an additional term, even in $p$, is introduced in the spinorial part of the Hamiltonian, $\vec h$,
which can be used to model, e.g., an external magnetic field: this term could also be considered in our framework but we preferred to neglect it
for the sake of simplicity.)
We remark that each of the Pauli components diffuses according to a classical drift-diffusion equation and, moreover, the spin has the additional
current term $4\alpha A_j(\vec n)$, coming from spin-orbit interactions, a relaxation term $- 4\alpha^2 B( \vec n)$, and an interaction with the external
potential, $-2\alpha \nabla^\perp V \times \vec n$, which shows the capability of controlling the spin by means of an applied voltage.
\section{Conclusions}
\label{sec:5}
In this paper, we have derived quantum drift diffusion equations (QDDE) for a 2DEG with spin-orbit interaction of Rashba type.
The derivation is based on the quantum version of the maximum entropy principle (Q-MEP), as proposed in Refs.\ \cite{DMR05,DR03}.
To our knowledge, this is the first application of the Q-MEP to the full spin structure and not only to the spin polarization (i.e.\
the projection of the spin vector on a given axis).

Our derivation starts with the formulation of a kinetic model which has an Hamiltonian part (basically, the mixed-state  Schr\"odinger equation
in the phase-space formulation) and a non-conservative, collisional term in the relaxation time approximation.
Here, the quantum equilibrium state given by the Q-MEP appears.

Assuming that the relaxation time is a small parameter in the problem, we apply the Chapman-Enskog technique to derive the quantum
drift-diffusion model \eqref{QDDE1}, \eqref{bkTg}, and \eqref{bkTTg}.
It forms a system of four equations: one for the charge density $n_0$ and three for the spin-vector components $\vec n = (n_1,n_2,n_3)$.
Such equations are non-local in the components $n_k$, since they are expressed in terms of Lagrange multipliers that are connected with the densities
by the (integral) constraint that the equilibrium state possesses such densities.
This aspect of the model is not different from the analogous QDDE obtained in the scalar \cite{DMR05,J2009} or bipolar  \cite{BM2010} cases.

A new feature of the present model is that the application of the Chapman-Enskog technique is not the standard one for the diffusive case
and resembles more to the Chapman-Enskog expansion of the hydrodynamic case.
This is due to the fact that, due to the peculiar form of the spin-orbit interaction, the equilibrium state has no zero current.
In the derivation, we have obtained a general condition, Eq.\ \eqref{commcond}, for such current to vanish.

Typically, the QDDE are expanded semiclassically, i.e.\ in powers of the scaled Planck constant $\eps$, which allows for an approximation of the QDDE
by a local model consisting in classical diffusive equations with ``quantum corrections''.
Here, we just computed the approximation at the leading order, in order to compare the semiclassical limit of our model with the semiclassical models already existing in the literature.
The semiclassical expansion of our QDDE, which is not an easy task, goes beyond the aim of the present paper and is deferred to a work in preparation.

\begin{acknowledgement}
The last two authors acknowledge partial support from
the Austrian Science Fund (FWF), grants F65, P30000, P33010, and W1245.
\end{acknowledgement}

\end{document}